\documentclass[envcountsame]{llncs}

\usepackage{makeidx}  
\usepackage{comment}

\newcommand{\rela}{\mathbf{A}}
\newcommand{\relb}{\mathbf{B}}
\newcommand{\relt}{\mathbf{T}}

\newcommand{\dom}{\mathsf{dom}}

\newcommand{\twk}{\mathsf{treewidth}<k}
\newcommand{\htwk}{\mathcal{H}(\mathsf{treewidth}<k)}

\newcommand{\cwk}{\mathsf{coverwidth} \leq k}
\newcommand{\hcwk}{\mathcal{H}(\mathsf{coverwidth} \leq k)}

\newcommand{\csp}{\mathsf{CSP}}
\newcommand{\qcsp}{\mathsf{QCSP}}

\begin{document}

\title{Beyond Hypertree Width: Decomposition Methods Without Decompositions}
\titlerunning{Beyond}

\author{Hubie Chen \and V\'{\i}ctor Dalmau}

\authorrunning{Hubie Chen and V\'{\i}ctor Dalmau}   

\institute{
Departament de Tecnologia\\
Universitat Pompeu Fabra\\
Barcelona, Spain\\
\email{\{hubie.chen,victor.dalmau\}@upf.edu}
}

\maketitle              

\begin{abstract}
The general intractability of the constraint satisfaction problem
has motivated the study of restrictions on this problem
that permit polynomial-time 
solvability.  
One major line of work has focused on structural restrictions,
which arise from restricting the interaction among constraint scopes.
In this paper, we engage in a mathematical investigation of
generalized hypertree width, a structural measure that has up to recently
eluded study.  We obtain a number of computational results,
including a simple proof of the tractability of CSP instances
having bounded generalized hypertree width.
\end{abstract}

\section{Introduction}

The \emph{constraint satisfaction problem} (CSP) is widely acknowledged
as a convenient framework for modelling search problems.
Instances of the CSP arise in a variety of domains, including
artificial intelligence, database theory, algebra, propositional logic,
and graph theory.
An instance of the CSP consists of a set of constraints on a 
set of variables;
the question is to determine if there is an assignment to  the
variables satisfying all of the constraints.
Alternatively, the CSP can be cast as the fundamental algebraic
problem of deciding, given two relational structures $\rela$ and $\relb$,
whether or not there is a homomorphism from $\rela$ to $\relb$.
In this formalization, each relation of $\rela$ contains the tuples of
variables that are constrained together, which are often called
the \emph{constraint scopes},
and the corresponding relation
of $\relb$ contains the allowable tuples of values that the
variable tuples may take.

It is well-known that the CSP, in its general formulation, is NP-complete;
this general intractability has motivated a large and rich
body of research aimed at identifying and understanding
restricted cases of the CSP that are polynomial-time tractable.
The restrictions that have been studied can, by and large,
be placed into one of two categories, which--due to the
homomorphism formulation of the CSP--have become known as
\emph{left-hand side} restrictions and \emph{right-hand side} restrictions.
From a high level view, left-hand side restrictions, also known as
\emph{structural restrictions},  
arise from prespecifying
a class of relational structures $\mathcal{A}$ from which the
left-hand side structure $\rela$ must come, while right-hand side
restrictions 
arise from prespecifying
a class of relational structures $\mathcal{B}$ from which the
right-hand side structure $\relb$ must come.
As this paper is concerned principally with structural restrictions,
we will not say more about right-hand side restrictions than that
their systematic study  has 
origins in a classic theorem of Schaefer~\cite{schaefer78},
and that recent years have seen some extremely exciting results on them
(for instance \cite{bulatov02-dichotomy,bulatov03-conservative}).

The structural restrictions studied in the literature
can all be phrased as restrictions on
the hypergraph $H(\rela)$ 
naturally arising from the left-hand side relational structure $\rela$, namely,
the hypergraph $H(\rela)$ with an edge $\{ a_1, \ldots, a_k \}$ for each
tuple $(a_1, \ldots, a_k)$ of $\rela$.
Let us briefly review some of the relevant results that have been obtained
on structural tractability.
The tractability of left-hand side relational structures having 
\emph{bounded treewidth}
was shown in the constraint satisfaction literature by
Dechter and Pearl \cite{dp89} and Freuder \cite{freuder90}.\footnote{
One way to define what we mean by treewidth here is the
treewidth of the graph obtained from $H(\rela)$ by drawing an edge
between any two vertices that are in the same hyperedge.}
Later, Dalmau et al. \cite{dkv02} building on ideas of Kolaitis and Vardi
\cite{kv00-conjunctive,kv00-game} gave a consistency-style
algorithm for deciding the bounded treewidth CSP.
For our present purposes, it is worth highlighting that although
the notion of bounded treewidth is defined in terms of
\emph{tree decompositions}, which can be computed efficiently
(under bounded treewidth), the algorithm given by Dalmau et al. \cite{dkv02}
does \emph{not} compute any form of tree decomposition.
Dalmau et al.
 also
identified a natural expansion
of structures having bounded treewidth 
that is tractable--namely,
the structures \emph{homomorphically equivalent}
to those having bounded treewidth.
The optimality of this latter result,
in the case of bounded arity,
was demonstrated by Grohe \cite{grohe03},
who proved--roughly speaking--that 
if the tuples of $\mathcal{A}$ are of bounded arity and
$\mathcal{A}$ gives rise to a tractable case of the CSP,
then it must fall into the natural expansion identified by
Dalmau et al. \cite{dkv02}. 

A number of papers, 
including \cite{gp84-hinges,gjc94-decomposing,gottlob00,gls01-acyclic,gls02-hypertree,cjg05}, 
have studied restrictions
that can be applied to relational structures of unbounded arity.
(Note that any class of relational structures of unbounded arity
cannot have bounded treewidth.)  
In a survey \cite{gottlob00}, Gottlob et al. 
show that the restriction of bounded hypertree width~\cite{gls02-hypertree}
is the most powerful structural restriction
for the CSP 
in that every other structural restriction studied in the literature
is subsumed by it.  Since this work \cite{gls02-hypertree,gottlob00},
whether or not there is a more general structural restriction than 
bounded hypertree width that ensures tractability,
has been a tantalizing open question.

In this paper, we study
 \emph{generalized hypertree width}, 
a structural measure for hypergraphs 
defined in \cite{gls03-characterizations} 
that is a natural variation of hypertree width;
we call this measure \emph{coverwidth}.
Coverwidth is \emph{trivially} upper-bounded by 
hypertree width, and so any class of hypergraphs having bounded
hypertree width has bounded coverwidth.
We define a combinatorial pebble game that can be played on any 
CSP instance, and demonstrate
that this game is intimately linked to coverwidth (Theorem~\ref{thm:main}).
Our study of coverwidth is conceptually simple,
mathematically elegant, and relatively compact; 
we believe that this hints that coverwidth is in fact a natural and robust
mathematical concept that may find further applications.
Overall, the investigation we perform takes
significant inspiration from 
methods, concepts, and ideas 
developed by
Kolaitis, Vardi, and coauthors \cite{kv00-conjunctive,kv00-game,dkv02,akv04}
that link together 
CSP consistency algorithms, 
the existential $k$-pebble games of Kolaitis and Vardi \cite{kv95-datalog},
and bounded treewidth.

Using the pebble game perspective, we are able to derive a number
of computational results.
One is that
the structural restriction of
\emph{bounded coverwidth} implies polynomial-time tractability; this
result
generalizes the tractability of bounded hypertree width.
It has been independently shown by Adler et al. that
the hypertree width of a hypergraph is 
linearly related to the coverwidth~\cite{agg}.
This result can be used in conjunction with the tractability of bounded
hypertree width to derive the tractability of bounded coverwidth.
However, we believe our proof of bounded coverwidth tractability to
be \emph{considerably simpler} than the
known proof 
of bounded hypertree width tractability~\cite{gls02-hypertree},
even though our proof is of a more general result.





To describe our results in greater detail, it will be useful to 
identify two computational problems that every form of structural
restriction gives rise to: a \emph{promise} problem, and
a \emph{no-promise} problem.  
In both problems, the goal is to identify all CSP instances 
obeying the structural restriction as either satisfiable or unsatisfiable.
In the promise problem, the input is a CSP instance 
that is \emph{guaranteed} to obey the structural restriction,
whereas in the no-promise problem, the input is an \emph{arbitrary} CSP
instance, and an algorithm may, on an instance not obeying the structural
restriction, decline to identify the instance as satisfiable or 
unsatisfiable.
Of course, CSPs arising in practice do \emph{not} come with guarantees
that they obey structural restrictions, 
and hence
an algorithm solving the no-promise problem is clearly the more desirable.
Notice that, for any structural restriction
having a polynomial-time solvable promise problem, if it is possible
to solve the \emph{identification problem} of 
deciding whether or not an instance obeys the restriction,
 in polynomial time,
then the no-promise problem is also polynomial-time solvable.
For bounded hypertree width, both the identification problem and
the no-promise problem are polynomial-time solvable.
In fact, the survey by Gottlob et al. \cite{gottlob00} only considers
structural restrictions for which the identification problem is
polynomial-time solvable, and thus only considers structural restrictions
for which the no-promise problem is polynomial-time solvable.

One of our main theorems (Theorem~\ref{thm:polytime}) is that 
the promise problem for bounded coverwidth is polynomial-time tractable,
via a general consistency-like algorithm.
In particular, we show that, on an instance having
bounded coverwidth, our algorithm detects an inconsistency if and only if
the instance is unsatisfiable.  
Our algorithm,
like the consistency algorithm of Dalmau et al. \cite{dkv02} for bounded
treewidth,  can be applied to \emph{any} CSP instance
to obtain a more constrained instance; 
our algorithm does \emph{not} need nor compute
any form of decomposition, even though the notion of coverwidth
is defined in terms of decompositions!

Intriguingly, we are then able to give a \emph{simple} algorithm for the
no-promise problem for bounded coverwidth 
(Theorem~\ref{thm:no-promise})
that 
employs 
 the consistency-like algorithm for the promise problem.
The behavior of this algorithm 
is reminiscent of self-reducibility arguments in
computational complexity theory, and on an instance of bounded coverwidth,
the algorithm is guaranteed to either report a satisfying
assignment or that the instance is unsatisfiable.
We believe that our results offer a direct challenge to the view
of structural tractability advanced in the Gottlob et al. survey
\cite{gottlob00}, since we are able to 
give a polynomial-time algorithm for the bounded coverwidth no-promise problem
without explicitly showing that there is a polynomial-time algorithm
for the bounded coverwidth identification problem.

Returning to the promise problem, we then show that
the tractability of bounded coverwidth structures can be generalized
to yield the tractability of structures 
\emph{homomorphically equivalent} to those having bounded coverwidth
(Theorem~\ref{thm:homom-equiv}).  This expansion of bounded coverwidth
tractability is analogous to the expansion of bounded treewidth tractability
carried out in \cite{dkv02}.

In the last section of this paper, we use the developed theory as well
as ideas in~\cite{cd05-qcsp} to define a tractable class of quantified
constraint satisfaction problems based on coverwidth.

We emphasize that \emph{none} of the algorithms in this paper need or compute
any type of decomposition, even though all of the structural restrictions
that they address are defined in terms of decompositions.

\paragraph{\bf Definitions.}
In this paper, we formalize the CSP as a relational homomorphism problem.
We review the relevant definitions that will be used.
A \emph{relational signature} is a finite set of relation symbols,
each of which has an associated arity.
A \emph{relational structure} $\rela$ (over signature $\sigma$)
consists of a universe $A$ and a relation $R^{\rela}$ over
$A$ for each relation symbol $R$ (of $\sigma$), such that the arity
of $R^{\rela}$ matches the arity associated to $R$.
We refer to the elements of the universe of a relational structure $\rela$
as $\rela$-elements.
When $\rela$ is a relational structure over $\sigma$ and
$R$ is any relation symbol of $\sigma$, the elements of
$R^{\rela}$ are called $\rela$-tuples.
Throughout this paper, we assume that all relational structures
under discussion have a finite universe.  We use boldface letters
$\rela, \relb, \ldots$ to denote relational structures.

A \emph{homomorphism} from a relational structure $\rela$ to another
relational structure $\relb$ is a mapping $h$ from the universe of
$\rela$ to the universe of $\relb$ such that for every relation symbol $R$
and every tuple $(a_1, \ldots, a_k) \in R^{\rela}$, it holds that
$(h(a_1), \ldots, h(a_k)) \in R^{\relb}$.  
(Here, $k$ denotes the arity of $R$.)
The \emph{constraint satisfaction problem (CSP)} is to decide,
given an ordered pair $\rela, \relb$ of relational structures, whether
or not there is a homomorphism from the first structure, $\rela$, to the 
second, $\relb$.  A homomorphism from $\rela$ to $\relb$ in an instance
$\rela, \relb$ of the CSP is also called a \emph{satisfying assignment},
and when a satisfying assignment exists, we will say that the instance
is \emph{satisfiable}.

\section{Coverwidth}

This section defines the structural measure of hypergraph complexity that
we call \emph{coverwidth}.  
As we have mentioned, coverwidth is equal to generalized hypertree width,
which was defined in~\cite{gls03-characterizations}.
We begin by defining the notion of 
\emph{hypergraph}.

\begin{definition}
A \emph{hypergraph} is an ordered pair $(V, E)$ consisting of 
a \emph{vertex set} $V$ and a \emph{hyperedge set} $E$.
The elements of $E$ are called \emph{hyperedges}; each
hyperedge is a subset of $V$.
\end{definition}

Basic to the measure of coverwidth is the notion of a tree decomposition.

\begin{definition}
A \emph{tree decomposition} of a hypergraph $(V, E)$
is a pair 
$$(T = (I, F), \{ X_i \}_{i \in I})$$
where 
\begin{itemize}
\item $T = (I, F)$ is a tree, and
\item each $X_i$ (with $i \in I$) is called a \emph{bag} and is a subset of $V$, 
\end{itemize}
such that
the following conditions hold:
\begin{enumerate}

\item $V = \cup_{i \in I} X_i$.
\item For all hyperedges $e \in E$, 
there exists $i \in I$ with $e \subseteq X_i$.
\item For all $v \in V$, the vertices $T_v = \{ i \in I: v \in X_i \}$
form a connected subtree of $T$.

\end{enumerate}
\end{definition}

Tree decompositions are generally applied to graphs, and in the context
of graphs, the measure of \emph{treewidth} has been heavily studied.
The \emph{treewidth} of a graph $G$ is the minimum of the quantity
$\max_{i \in I} |X_i| - 1$ over all tree decompositions of $G$.
In other words, a tree decomposition is measured based on 
 its largest bag, and the treewidth is then defined
based on the ``lowest cost'' tree decomposition.

The measure of coverwidth is also based
on the notion of tree decomposition.  In coverwidth, 
a tree decomposition is also measured based on its ``largest'' bag;
however, the measure applied to a bag is the number of hyperedges
needed to cover it, called here the \emph{weight}.

\begin{definition}
A $k$-union over a hypergraph $H$ 
(with $k \geq 0$)
is the union $e_1 \cup \ldots \cup e_k$
of $k$ edges $e_1, \ldots, e_k$ of $H$.
\end{definition}

The empty set is considered to be the unique $0$-union over a hypergraph.

\begin{definition}
\label{def:weight-subset}
Let $H = (V, E)$ be a hypergraph.
The \emph{weight} of a subset $X \subseteq V$
is the smallest integer $k \geq 0$ such that
$X \cap (\cup_{e \in E} e)$ is contained in a $k$-union over $H$.
\end{definition}

We measure a tree decomposition according to its heaviest bag,
and define the \emph{coverwidth} of a hypergraph according to the
lightest-weight tree decomposition.

\begin{definition}
The \emph{weight} of a tree decomposition of $H$ is the maximum weight
over all of its bags.
\end{definition}

\begin{definition}
The \emph{coverwidth} of a hypergraph $H$ is the minimum weight
over all tree decompositions of $H$.
\end{definition}

It is straightforward to verify that the coverwidth of a hypergraph
is equal to the generalized hypertree width 
of a hypergraph~\cite{gls03-characterizations}.
Since the generalized hypertree width of a hypergraph is always
less than or equal to its hypertree width,
coverwidth is at least
as strong as hypertree width in that results on bounded coverwidth
imply results on bounded hypertree width.

There is another formulation of tree decompositions 
 that is
often wieldy, see for instance \cite{bod05}.

\begin{definition}
\label{def:scheme}
A \emph{scheme} of a hypergraph $H = (V, E)$
is a graph $(V, F)$ such that
\begin{itemize}
\item $(V,F)$ has a perfect elimination ordering, that is, an ordering 
$v_1, \ldots, v_n$
of its vertices such that for all $i<j<k$, if $(v_i,v_k)$ and $(v_j,v_k)$
are edges in $F$, then $(v_i,v_j)$ is also an edge in $F$, and
\item the vertices of every hyperedge of $E$ induce a clique in $(V,F)$.
\end{itemize}
\end{definition}

It is well known that the property of having a perfect elimination ordering is equivalent to being chordal.
The following proposition is also well-known.

\begin{proposition}
\label{prop:decomposition-scheme}
Let $H$ be a hypergraph.
For every tree decomposition of $H$, there exists a scheme such that
each clique of the scheme is contained in a bag of the tree decomposition.
Likewise, for every scheme of $H$, there exists a tree decomposition such that
each bag of the tree decomposition is contained in a clique of the scheme.
\end{proposition}

Let us define the \emph{weight} of a scheme (of a hypergraph $H$)
to be the maximum weight (with respect to $H$) over all of its cliques.
The following proposition
is immediate from Proposition \ref{prop:decomposition-scheme}
and the definition of coverwidth, and 
can be taken as an alternative definition of coverwidth.

\begin{proposition}
\label{prop:coverwidth-schemes}
The coverwidth of a hypergraph $H$ is equal to the minimum weight
over all schemes of $H$.
\end{proposition}

We now define the hypergraph associated to a relational structure.
Roughly speaking, this hypergraph is obtained by ``forgetting'' the
ordering of the $\rela$-tuples.

\begin{definition}
Let $\rela$ be a relational structure.
The hypergraph associated to $\rela$ is denoted by $H(\rela)$;
the vertex set of $H(\rela)$ is the universe of $\rela$,
 and for each
$\rela$-tuple $(a_1, \ldots, a_k)$, there is an edge
$\{ a_1, \ldots, a_k \}$ in $H(\rela)$.
\end{definition}

We will often implicitly pass from a relational structure to its
associated hypergraph, that is, we simply write $\rela$ in place of $H(\rela)$.
In particular, we will speak of $k$-unions over a relational structure
$\rela$.

\section{Pebble Games}

We now define a class of pebble games for studying the measure of
coverwidth.  These games are essentially equivalent to the 
existential $k$-pebble games defined by 
Kolaitis and Vardi and used to study 
constraint satisfaction~\cite{kv95-datalog,kv00-game}.
The pebble game that we use is defined as follows.  The game is played
between two players, the \emph{Spoiler} and the \emph{Duplicator},
on a pair of relational structures $\rela, \relb$ that are defined
over the same signature.
Game play proceeds in rounds, and in each round one of the following
occurs:
\begin{enumerate}
\item The Spoiler places a pebble on an $\rela$-element $a$.
In this case, the Duplicator must respond by placing a corresponding pebble,
denoted by $h(a)$, on a $\relb$-element.

\item The Spoiler removes a pebble from an $\rela$-element $a$.
In this case, the corresponding pebble $h(a)$ on $\relb$ is removed.
\end{enumerate}

When game play begins, there are no pebbles on any $\rela$-elements,
nor on any $\relb$-elements, and so the first round is of the first type.
We assume that the Spoiler never places two pebbles on the same
$\rela$-element, so that $h$ is a partial function (as opposed to a relation).
The Duplicator wins the game if he can always ensure
that $h$ is a \emph{projective homomorphism} from $\rela$ to $\relb$;
otherwise, the Spoiler wins.
A \emph{projective homomorphism} (from $\rela$ to $\relb$)
is
a partial function $h$ from the universe of $\rela$ to the universe of $\relb$
such that
for any relation symbol $R$ and any tuple
$(a_1, \ldots, a_k) \in R^{\rela}$ of $\rela$,
there exists a tuple $(b_1, \ldots, b^k) \in R^{\relb}$
where $h(a_i) = b_i$ for all $a_i$ on which $h$ is defined.

As we mentioned, the definition of this game is based on the
\emph{existential $k$-pebble game} introduced by 
Kolaitis and Vardi~\cite{kv95-datalog,kv00-game}.
In the \emph{existential $k$-pebble game}, 
 the number of pebbles
that the Spoiler may use is bounded by $k$, 
and the Duplicator need only
must ensure that $h$ is a \emph{partial homomorphism}.
A close relationship
between this game and bounded treewidth has been identified~\cite{akv04}.

\begin{theorem} \cite{akv04}
\label{thm:game-treewidth}
Let $\rela$ and $\relb$ be relational structures.
For all $k \geq 2$, the following are equivalent.
\begin{itemize}

\item There is a winning strategy for the Duplicator
in the existential $k$-pebble game on $\rela, \relb$.

\item For all relational structures $\relt$ of treewidth $< k$,
if there is a homomorphism from $\relt$ to $\rela$, 
then there is a homomorphism from $\relt$ to $\relb$.
\end{itemize}
\end{theorem}

To relate the game that we have defined with coverwidth, 
we are interested in parameterized versions of the game where the
\emph{weight} of the pebbles that the Spoiler has in play, is bounded
by a constant $k$.
(Here, by ``weight'' we are using Definition~\ref{def:weight-subset}.)
That is, the weight of the
$\rela$-elements that have pebbles, is bounded by the constant $k$.
We call this the \emph{existential $k$-cover game}.
We now formalize the notion of a
\emph{winning strategy} for the Duplicator in the existential $k$-cover game.
Note that when $h$ is a partial function, we use $\dom(h)$ 
to denote the domain of $h$.

\begin{definition}
A \emph{winning strategy} for the Duplicator in the existential $k$-cover game
on relational structures $\rela, \relb$ 
is a non-empty set $H$ of projective homomorphisms
(from $\rela$ to $\relb$) having the following two properties.
\begin{enumerate}

\item (the ``forth'' property)
For every $h \in H$ and $\rela$-element $a \notin \dom(h)$,
if $\dom(h) \cup \{ a \}$ has weight $\leq k$, then there exists
a projective homomorphism $h' \in H$ extending $h$ with
$\dom(h') = \dom(h) \cup \{ a \}$.

\item The set $H$ is closed under subfunctions, that is, if $h \in H$
and $h$ extends $h'$, then $h' \in H$.

\end{enumerate}
\end{definition}

We have the following analog of Theorem \ref{thm:game-treewidth}.

\begin{theorem}
\label{thm:main}
Let $\rela$ and $\relb$ be relational structures.
For all $k \geq 1$, the following are equivalent.
\begin{itemize}

\item There is a winning strategy for the Duplicator
in the $k$-cover game on $\rela, \relb$.

\item For all relational structures $\relt$ of coverwidth $\leq k$,
if there is a homomorphism from $\relt$ to $\rela$, 
then there is a homomorphism from $\relt$ to $\relb$.
\end{itemize}
\end{theorem}

Theorem~\ref{thm:main} can be easily applied to show that 
in an instance $\rela, \relb$ of the CSP, if the left-hand side
structure has coverwidth bounded by $k$, 
then deciding if there is a homomorphism from $\rela$ to $\relb$
is equivalent to deciding
the existence of a Duplicator winning strategy
in the existential $k$-cover game.

\begin{theorem}
\label{thm:winning-strategy-homomorphism}
Let $\rela$ be a relational structure having coverwidth $\leq k$, and
let $\relb$ be an arbitrary relational structure.
There is a winning strategy for the Duplicator
in the $k$-cover game on $\rela, \relb$ if and only if
there is a homomorphism from $\rela$ to $\relb$.
\end{theorem}

We will use this theorem in the next section to develop tractability
results.  Although we use Theorem~\ref{thm:main} to derive this theorem,
we would like to emphasize that the full power of Theorem~\ref{thm:main}
is not needed to derive it, as pointed out in the proof.

\begin{proof}
If there is a homomorphism from $\rela$ to $\relb$, the Duplicator
can win by always setting pebbles according the homomorphism.
The other direction is immediate from Theorem~\ref{thm:main}
(note that we only need the forward implication and $\relt = \rela$).
\qed \end{proof}

\section{The Algorithmic Viewpoint}

The previous section introduced the \emph{existential $k$-cover game}.
We showed that deciding a CSP instance of bounded coverwidth is equivalent
to deciding if the Duplicator has a winning strategy in the
existential $k$-cover game.
In this section, we show that the latter property--the existence
of a Duplicator winning strategy--can be decided algorithmically in
polynomial time.  To this end, it will be helpful to introduce
the notion of a \emph{compact winning strategy}.

\begin{definition}
A \emph{compact winning strategy} 
for the Duplicator in the existential $k$-cover game
on relational structures $\rela, \relb$ 
is a non-empty set $H$ of projective homomorphisms
(from $\rela$ to $\relb$) having the following properties.
\begin{enumerate}

\item For all $h \in H$, $\dom(h)$ is a $k$-union (over $\rela$).

\item For every $h \in H$ and for every $k$-union $U$ (over $\rela$),
there exists $h' \in H$ with $\dom(h') = U$ such that for every 
$v \in \dom(h) \cap \dom(h')$, $h(v) = h'(v)$.

\end{enumerate}
\end{definition}

\begin{proposition}
\label{prop:compact}
In the existential $k$-cover game
on a pair of relational structures $\rela, \relb$,
the Duplicator has a winning strategy if and only if
the Duplicator has a compact winning strategy.
\end{proposition}

\begin{proof}
Suppose that the Duplicator has a winning strategy $H$.
Let $C$ be the set containing all functions $h \in H$ such that
$\dom(h)$ is a $k$-union.  We claim that $C$ is a compact winning strategy.
Clearly $C$ satisfies the first property of a compact winning strategy,
so we show that it satisfies the second property.
Suppose $h \in C$ and let $U$ be a $k$-union.
By the subfunction property of a winning strategy, the restriction
$r$ of $h$ to $\dom(h) \cap U$ is in $H$.  By repeated application of
the forth property, there is an extension $e$ of $r$ that is in $H$
and has domain $U$, which serves as the desired $h'$.

Now suppose that the Duplicator has a compact winning strategy $C$.
Let $H$ be the closure of $C$ under subfunctions.  We claim that
$H$ is a winning strategy.  It suffices to show that $H$ has the 
forth property.  Let $h \in H$ and suppose that $a$ is an $\rela$-element
where $\dom(h) \cup \{ a \}$ has weight $\leq k$.
Let $U$ be a $k$-union such that $\dom(h) \cup \{ a \} \subseteq U$.
By definition of $H$, there is a function $e \in C$ extending $h$.
Apply the second property of a compact winning strategy to $e$
and $U$ to obtain an $e' \in C$ with domain $U$ such that for every 
$v \in \dom(e) \cap \dom(e')$, $e(v) = e'(v)$.
Notice that $\dom(h) \subseteq \dom(e) \cap \dom(e')$.
Thus, the restriction of $e'$ to $\dom(h) \cup \{ a \}$ is in $H$ and extends
$h$.
\qed \end{proof}

We have just shown that deciding if there is a winning strategy,
in an instance of the existential $k$-cover game, is equivalent to
deciding if there is a compact winning strategy.  We now use this equivalence
to give a polynomial-time algorithm for deciding if there is a winning
strategy.

\begin{theorem}
\label{thm:algorithm-winning-strategy}
For all $k \geq 1$, there exists a polynomial-time algorithm that,
given a pair of relational structures $\rela, \relb$,
decides whether or not there is a winning strategy
for the Duplicator in the existential $k$-cover game on $\rela, \relb$.
\end{theorem}

\begin{proof}
By Proposition~\ref{prop:compact}, it suffices to give a polynomial-time
algorithm that decides if there is a compact winning strategy.
It is straightforward to develop such an algorithm based on the definition
of compact winning strategy.  Let $H$ be the set of all functions
$h$ such that $\dom(h)$ is a $k$-union (over $\rela$) and 
such that $h$ is a projective homomorphism from $\rela$ to $\relb$.
Iteratively perform the following until no changes can be made to $H$:
for every function $h \in H$ and every $k$-union $U$, check to see if there
is $h' \in H$ such that the second property (of compact winning strategy)
is satisfied; if not, remove $h$ from $H$.
Throughout the algorithm, we have maintained the invariant that 
any compact winning strategy must be a subset of $H$.
Hence, if when the algorithm terminates $H$ is empty, then there is no
compact winning strategy.  And if $H$ is non-empty when the algorithm
terminates, $H$ is clearly a compact winning strategy.

The number of $k$-unions (over $\rela$) is polynomial in the number of
tuples in $\rela$.  Also, for each $k$-union $U$, the number of projective
homomorphisms $h$ with $\dom(h) = U$ from $\rela$ to $\relb$ is
polynomial in the number of tuples in $\relb$.
Hence, the size of the original set $H$ is polynomial in the original
instance.  Since in each iteration an element is removed from $H$,
the algorithm terminates in polynomial time.
\qed \end{proof}

The algorithm we have just described 
in the proof of Theorem~\ref{thm:algorithm-winning-strategy}
may appear to be quite specialized.  However, we now show that
essentially that algorithm can be viewed as a \emph{general} inference
procedure for CSP instances in the vein of existing consistency algorithms.
In particular, we give a general
algorithm called \emph{projective $k$-consistency} for CSP instances
that, given a CSP instance, performs inference and 
outputs a more constrained CSP instance having
exactly the same satisfying assignments as the original.
On a CSP instance 
$\rela, \relb$, the algorithm might
detect an \emph{inconsistency}, by which we mean
that it detects that there is no 
homomorphism from $\rela$ to $\relb$.  If it does not, then it is 
guaranteed
that there is a winning strategy for the Duplicator.

\begin{definition}
\label{def:consistency-algorithm}
The \emph{projective $k$-consistency algorithm} takes as input
a CSP instance $\rela, \relb$, and consists of the following steps.
\begin{itemize}

\item Create a new CSP instance $\rela', \relb'$ as follows.
Let the universe of $\rela'$ be the universe of $\rela$,
and the universe of $\relb'$ be the universe of $\relb$.
Let the signature of $\rela'$ and $\relb'$ contain a relation
symbol $R_U$ for each $k$-union $U$ over $\rela$.
For each $k$-union $U$, the relation
$R_U^{\rela'}$ is defined as $(u_1, \ldots, u_k)$, where
$u_1, \ldots, u_k$ are exactly the elements of $U$ in some order;
and
$R_U^{\relb'}$ is defined as the set of all tuples
$(b_1, \ldots, b_k)$ such that the mapping taking
$u_i \rightarrow b_i$ is a projective homomorphism from $\rela$ to $\relb$.

\item Iteratively perform the following
until no changes can be made: remove any $\relb'$-tuple
$(b_1, \ldots, b_k)$ that is not a projective homomorphism.
We say that a $\relb'$-tuple $(b_1, \ldots, b_k) \in R_U^{\relb'}$
is a projective homomorphism if,
letting $(u_1, \ldots, u_k)$ denote the unique element of $R_U^{\rela'}$,
 the function taking $u_i \rightarrow b_i$ is a projective homomorphism
from $\rela'$ to $\relb'$.

\item Report an inconsistency if there are no $\relb'$-tuples
remaining.

\end{itemize}
\end{definition}

\begin{theorem}
\label{thm:consistency-algorithm}
For each $k \geq 1$, the projective $k$-consistency algorithm, given
as input a CSP instance $\rela, \relb$:

\begin{itemize}

\item runs in polynomial time,

\item outputs a CSP instance 
$\rela', \relb'$ that 
has the same satisfying assignments as $\rela, \relb$, and

\item reports an inconsistency if and only if the Duplicator
does not have a winning strategy in the existential $k$-cover game on
$\rela, \relb$.

\end{itemize}
\end{theorem}

\begin{proof}
The first property is straightforward to verify.  For the second property,
 observe that, each time a tuple is removed from $\relb'$, 
the set of satisfying assignments is preserved.  For the third property,
observe that, associating $\relb'$-tuples to functions as in
Definition~\ref{def:consistency-algorithm}, the behavior of 
the projective $k$-consistency algorithm is identical to the behavior of
the algorithm in the proof of Proposition \ref{prop:compact}.
\qed \end{proof}

By using the results presented in this section thus far, it is easy
to show that CSP instances of bounded coverwidth are tractable.
Define the coverwidth of a CSP instance $\rela, \relb$ to be the
coverwidth of $\rela$.
Let $\csp[\cwk]$ be the restriction of the CSP to all instances
of coverwidth less than or equal to $k$.

\begin{theorem}
\label{thm:polytime}
For all $k \geq 1$, the problem $\csp[\cwk]$ is decidable in polynomial time
by the projective $k$-consistency algorithm.
In particular, on an instance of $\csp[\cwk]$, the 
projective $k$-consistency algorithm reports an inconsistency
if and only if the instance is not satisfiable.
\end{theorem}

\begin{proof}
Immediate from Theorem \ref{thm:winning-strategy-homomorphism} 
and the third property of Theorem \ref{thm:consistency-algorithm}.
\qed \end{proof}

Note that we can derive the tractability of CSP instances having
bounded hypertree width immediately from 
Theorem~\ref{thm:polytime}.

Now, given a CSP instance that is promised to have bounded coverwidth,
we can use projective $k$-consistency to decide the instance
(Theorem \ref{thm:polytime}).  This tractability result
can in fact be pushed further: we can show that 
there is a generic polynomial-time that, given
an \emph{arbitrary} CSP instance, 
 is \emph{guaranteed} to decide instances of bounded coverwidth.
Moreover, whenever an instance is decided to be a ``yes'' instance
by the algorithm,
a satisfying assignment is constructed.

\begin{theorem}
\label{thm:no-promise}
For all $k \geq 1$, there exists a polynomial-time algorithm that,
given any CSP instance $\rela, \relb$,
\begin{enumerate}
\item outputs a satisfying assignment for $\rela, \relb$,
\item correctly reports that $\rela, \relb$ is unsatisfiable, or
\item reports ``I don't know''.
\end{enumerate}
The algorithm always performs (1) or (2) on an instance of $\csp[\cwk]$.
\end{theorem}

\begin{proof}
The algorithm is a simple extension of the projective $k$-consistency
algorithm.  First, the algorithm applies the projective $k$-consistency
algorithm; if an inconsistency is detected, then the algorithm terminates
and reports that $\rela, \relb$ is unsatisfiable.
Otherwise, it initializes
$V$ to be the universe $A$ of $\rela$, and does the following:
\begin{itemize}
\item If $V$ is empty, terminate and identify the mapping taking
each $a \in A$ to the $\relb$-element in $R_a^{\relb}$, as
a satisfying assignment.
\item Pick any variable $v \in V$.
\item Expand the signature of $\rela, \relb$ to include another symbol
$R_v$ with $R_v^{\rela} = \{ (v) \}$.
\item Try to find a $\relb$-element $b$ such that when
$R_v^{\relb}$ is set to $\{ (b) \}$, no inconsistency is detected by
the projective $k$-consistency algorithm on the expanded instance.
\begin{itemize}

\item If there is no such $\relb$-element, terminate and report
``I don't know''.

\item Otherwise, set $R_v^{\relb}$ to such a $\relb$-element,
remove $v$ from $V$, and repeat from the first step using the expanded
instance.
\end{itemize}
\end{itemize}
If the procedure terminates from $V$ being empty in the first step,
the mapping that is output is
straightforwardly verified to be a satisfying assignment.

Suppose that the algorithm is given an instance of $\csp[\cwk]$.
If it is unsatisfiable, then the algorithm reports that the instance
is unsatisfiable by Theorem \ref{thm:polytime}.
So suppose that the instance is satisfiable.  
We claim that each iteration
preserves the satisfiability of the instance.
Let $\rela, \relb$ denote the CSP instance at the beginning of 
an arbitrary iteration of the algorithm.
If no inconsistency is detected after
adding a new relation symbol $R_v$ with
$R_v^{\rela} = \{ (v) \}$ and
$R_v^{\relb} = \{ (b) \}$, there must be a satisfying assignment
mapping $v$ to $b$ by Theorem~\ref{thm:polytime}.
Note that adding unary relation symbols to a CSP instance
does not change the coverwidth of the instance.
\qed \end{proof}

We now expand the tractability result of Theorem \ref{thm:polytime}, 
and show
the tractability of CSP instances that are
\emph{homomorphically equivalent} to instances
of bounded coverwidth.
Formally, let us say that $\rela$ and $\rela'$ are homomorphically
equivalent if there is a homomorphism from $\rela$ to $\rela'$
as well as a homomorphism from $\rela'$ to $\rela$.
Let $\csp[\hcwk]$ denote the restriction of the CSP to instances
$\rela, \relb$ where $\rela$ is homomorphically equivalent to
a relational structure of coverwidth less than or equal to $k$.

\begin{theorem}
\label{thm:homom-equiv}
For all $k \geq 1$, the problem $\csp[\hcwk]$ is decidable in polynomial time
by the projective $k$-consistency algorithm.
In particular, on an instance of $\csp[\hcwk]$, the 
projective $k$-consistency algorithm reports an inconsistency
if and only if the instance is not satisfiable.
\end{theorem}


\begin{proof}
Let $\rela, \relb$ be a CSP instance where $\rela$
is homomorphically equivalent to a relational structure
$\rela'$ of coverwidth $\leq k$.
The following conditions are equivalent; after stating each condition,
we indicate how to show equivalence with the previous condition.
\begin{itemize}
\item There is a homomorphism from $\rela$ to $\relb$.
\item There is a homomorphism from $\rela'$ to $\relb$ (straightforward).
\item The Duplicator has a winning strategy in the existential $k$-cover game on $\rela', \relb$ 
(Theorem~\ref{thm:winning-strategy-homomorphism}).
\item The Duplicator has a winning strategy in the existential $k$-cover game on $\rela, \relb$
(Theorem~\ref{thm:main}).
\item The projective $k$-consistency algorithm does not report an inconsistency on $\rela, \relb$ (Theorem~\ref{thm:consistency-algorithm}).
\end{itemize}
\qed \end{proof}

\section{Quantified Constraint Satisfaction}

We now sketch how the ideas given in this paper on constraint satisfaction
can be combined with the ideas in \cite{cd05-qcsp} to yield results
on quantified constraint satisfaction.  Specifically, we define a notion
of coverwidth for QCSPs, and show that bounded coverwidth QCSPs are
tractable.

\paragraph{\bf Definitions.}
We first briefly define the QCSP and relevant associated notions.
A \emph{quantified relational structure} is a pair
$(p, \rela)$ where $\rela$ is a relational structure
and $p$ is a \emph{quantifier prefix}, an expression of the form 
$Q_1 v_1 \ldots Q_n v_n$ where 
each $Q_i$ is a quantifier (either $\exists$ or $\forall$) and 
$v_1, \ldots, v_n$ are exactly the elements of the universe of $\rela$.
The quantified constraint formula $\phi_{(p, \rela)}$
associated to a quantified relational structure
$(p, \rela)$ (where $\rela$ is over signature $\sigma$)
is defined to be the formula $p  \mathcal{C}_{\rela}$,
where $\mathcal{C}_{\rela}$ is the conjunction of all atomic formulas
in the set 
$\{ R(a_1, \ldots, a_k): R \in \sigma, (a_1, \ldots, a_k) \in R^{\rela} \}$.
We say that there is a homomorphism from $(p, \rela)$ to $\relb$
if $\relb \models \phi_{(p, \rela)}$.
We define the QCSP as the problem of deciding, given a
quantified relational structure $(p, \rela)$ and a relational structure
$\relb$, if there is a homomorphism from $(p, \rela)$ to $\relb$.

A quantifier prefix $p = Q_1 v_1 \ldots Q_n v_n$ 
can be viewed as the concatenation
of \emph{quantifier blocks} where quantifiers in each block are the same,
and consecutive quantifier blocks have different quantifiers.
For example, the quantifier prefix
$\forall v_1 \forall v_2 \exists v_3 \forall v_4 \forall v_5 \exists v_6 \exists v_7 \exists v_8$, 
consists of four quantifier blocks:
$\forall v_1 \forall v_2$,
$\exists v_3$,
$\forall v_4 \forall v_5$, and
$\exists v_6 \exists v_7 \exists v_8$.  
We say that a variable $v_j$ \emph{comes after} a variable $v_i$ in $p$
if they are in the same quantifier block, or
$v_j$ is in a
quantifier block following the quantifier block of $v_i$.
Equivalently, the variable $v_j$ comes after the variable $v_i$ in $p$
if one of the following conditions holds: (1) $j \geq i$, or
(2) $j < i$ and all of the quantifiers $Q_j, \ldots, Q_i$ are of the same
type.  

\paragraph{\bf Coverwidth.}
We now define a notion of coverwidth for quantified relational structures.
This can be viewed as a generalization of the definition of coverwidth
in terms of schemes, given in Proposition~\ref{prop:coverwidth-schemes}.

\begin{definition}
A \emph{scheme} of a quantified relational structure $(p, \rela)$
is a scheme $(V, F)$ of the hypergraph $H(\rela)$
(in the sense of Definition \ref{prop:coverwidth-schemes})
such that $(V, F)$ has a perfect elimination ordering 
$v_1, \ldots, v_n$
respecting
the quantifier prefix $p$ in that if $i < j$, then $v_j$ comes after $v_i$
in $p$.
\end{definition}

\begin{definition}
The coverwidth of a quantified relational structure $(p, \rela)$
is equal to the minimum weight (with respect to $\rela$)
over all schemes of $(p, \rela)$.
\end{definition}

\paragraph{\bf The quantified $k$-cover game.}
We can naturally extend the
$k$-cover game, making use of ideas from \cite{cd05-qcsp}, to
define the \emph{quantified $k$-cover game}.  
We describe the quantified $k$-cover game
 by defining the notion of a winning strategy
for the Duplicator.

\begin{definition}
A \emph{winning strategy} for the Duplicator in the quantified $k$-cover game
on $(p, \rela)$ and $\relb$ is a non-empty set $H$ of projective homomorphisms
(from $\rela$ to $\relb$) having the following properties.
\begin{enumerate}

\item For every $h \in H$ and every existentially quantified $\rela$-element
$a \notin \dom(h)$ coming after all elements of $\dom(h)$,
if $\dom(h) \cup \{ a \}$ has weight $\leq k$, then 
there exists a projective homomorphism $h' \in H$ extending $h$ with
$\dom(h') = \dom(h) \cup \{ a \}$.

\item For every $h \in H$,
every $\relb$-element $b$, 
and every universally quantified $\rela$-element
$a \notin \dom(h)$ coming after all elements of $\dom(h)$,
if $\dom(h) \cup \{ a \}$ has weight $\leq k$, then 
there exists a projective homomorphism $h' \in H$ extending $h$ with
$\dom(h') = \dom(h) \cup \{ a \}$ and $h'(a) = b$.

\item The set $H$ is closed under subfunctions, that is, if $h \in H$
and $h$ extends $h'$, then $h' \in H$.

\end{enumerate}
\end{definition}

We have the following analog of 
Theorem~\ref{thm:winning-strategy-homomorphism}.

\begin{theorem}
Let $(p, \rela)$ be a 
quantified relational structure having coverwidth $\leq k$, and
let $\relb$ be an arbitrary relational structure.
There is a winning strategy for the Duplicator
in the quantified $k$-cover game on $(p, \rela), \relb$ if and only if
there is a homomorphism from $(p, \rela)$ to $\relb$.
\end{theorem}

\paragraph{\bf Tractability.}
We let $\qcsp[\cwk]$ denote 
the restriction of the QCSP to all instances
$(p, \rela), \relb$ where $(p, \rela)$ 
has coverwidth less than or equal to $k$.
We have the following tractability result.

\begin{theorem}
\label{thm:coverwidth-qcsp}
For all $k \geq 1$, the problem $\qcsp[\cwk]$ is decidable in polynomial time.
\end{theorem}

The algorithm for Theorem~\ref{thm:coverwidth-qcsp} is similar to projective
$k$-consistency, but in addition to removing tuples that are not
projective homomorphisms, it removes further tuples, as follows.
Let $A_1 \ldots A_m$ denote the quantifier blocks of the prefix $p$,
and let $A_i$ be an existential quantifier block.
Take a tuple from the right-hand side structure, 
view it as a mapping $h: U \rightarrow B$, and consider its restriction
$h'$ to
$A_1 \cup \ldots \cup A_i$.
Let $Y$ be the set of all universally quantified variables
in $U \cap (A_{i+1} \cup \ldots \cup A_m)$.  If there exists any
extension of $h'$ to $\dom(h') \cup Y$ that is \emph{not} a projective
homomorphism, then the tuple is removed.
After the procedure terminates, if no inconsistency is detected,
then the projective homomorphisms $h$ of the new instance where the
weight of $\dom(h)$ is $\leq k$, is a
winning strategy for the Duplicator in the quantified $k$-cover game
on the original instance.

We can expand the result of
Theorem \ref{thm:coverwidth-qcsp}
 by Q-homomorphic equivalence, defined in
\cite{cd05-qcsp}.  Let $\qcsp[\hcwk]$ denote 
the restriction of the QCSP to all instances
$(p, \rela), \relb$ where $(p, \rela)$ is Q-homomorphically equivalent
to a quantified relational structure that
has coverwidth less than or equal to $k$.

\begin{theorem}
For all $k \geq 1$, the problem $\qcsp[\hcwk]$ is decidable in polynomial time.
\end{theorem}

\bibliographystyle{plain}
\bibliography{csp}

\begin{thebibliography}{10}

\bibitem{agg}
Isolde Adler, Georg Gottlob, and Martin Grohe.
\newblock Hypertree-width and related hypergraph invariants.
\newblock In preparation.

\bibitem{akv04}
A.~Atserias, Ph.~G. Kolaitis, and M.~Y. Vardi.
\newblock Constraint propagation as a proof system.
\newblock In {\em CP 2004}, 2004.

\bibitem{bod05}
Hans~L. Bodlaender.
\newblock Discovering treewidth.
\newblock In {\em {SOFSEM} 2005}, 2005.

\bibitem{bulatov02-dichotomy}
Andrei Bulatov.
\newblock A dichotomy theorem for constraints on a three-element set.
\newblock In {\em Proceedings of 43rd {IEEE Symposium on Foundations of
  Computer Science}}, pages 649--658, 2002.

\bibitem{bulatov03-conservative}
Andrei Bulatov.
\newblock Tractable conservative constraint satisfaction problems.
\newblock In {\em Proceedings of 18th {IEEE Symposium on Logic in Computer
  Science (LICS '03)}}, pages 321--330, 2003.
\newblock Extended version appears as Oxford University technical report
  {PRG-RR--03-01.}

\bibitem{cd05-qcsp}
Hubie Chen and Victor Dalmau.
\newblock From pebble games to tractability: An ambidextrous consistency
  algorithm for quantified constraint satisfaction.
\newblock Manuscript, 2005.

\bibitem{cjg05}
D.~Cohen, P.~Jeavons, and M.~Gyssens.
\newblock A unified theory of structural tractability for constraint
  satisfaction and spread cut decomposition.
\newblock To appear in IJCAI 2005, 2005.

\bibitem{dkv02}
Victor Dalmau, Phokion~G. Kolaitis, and Moshe~Y. Vardi.
\newblock Constraint satisfaction, bounded treewidth, and finite-variable
  logics.
\newblock In {\em Constraint Programming '02}, LNCS, 2002.

\bibitem{dp89}
Rina Dechter and Judea Pearl.
\newblock Tree clustering for constraint networks.
\newblock {\em Artificial Intelligence}, pages 353--366, 1989.

\bibitem{freuder90}
Eugene Freuder.
\newblock Complexity of $k$-tree structured constraint satisfaction problems.
\newblock In {\em AAAI-90}, 1990.

\bibitem{gls02-hypertree}
G.~Gottlob, L.~Leone, and F.~Scarcello.
\newblock Hypertree decomposition and tractable queries.
\newblock {\em Journal of Computer and System Sciences}, 64(3):579--627, 2002.

\bibitem{gls03-characterizations}
G.~Gottlob, L.~Leone, and F.~Scarcello.
\newblock Robbers, marshals, and guards: game theoretic and logical
  characterizations of hypertree width.
\newblock {\em Journal of Computer and System Sciences}, 66:775--808, 2003.

\bibitem{gottlob00}
Georg Gottlob, Nicola Leone, and Francesco Scarcello.
\newblock A comparison of structural csp decomposition methods.
\newblock {\em Artif. Intell.}, 124(2):243--282, 2000.

\bibitem{gls01-acyclic}
Georg Gottlob, Nicola Leone, and Francesco Scarcello.
\newblock The complexity of acyclic conjunctive queries.
\newblock {\em Journal of the {ACM}}, 43(3):431--498, 2001.

\bibitem{grohe03}
Martin Grohe.
\newblock The complexity of homomorphism and constraint satisfaction problems
  seen from the other side.
\newblock In {\em FOCS 2003}, pages 552--561, 2003.

\bibitem{gjc94-decomposing}
M.~Gyssens, P.G. Jeavons, and D.A. Cohen.
\newblock Decomposing constraint satisfaction problems using database
  techniques.
\newblock {\em Artificial Intelligence}, 66(1):57--89, 1994.

\bibitem{gp84-hinges}
M.~Gysssens and J.~Paradaens.
\newblock A decomposition methodology for cyclic databases.
\newblock In {\em Advances in Database Theory}, volume~2, pages 85--122. Plenum
  Press, New York, NY, 1984.

\bibitem{kv95-datalog}
Ph.G. Kolaitis and M.Y. Vardi.
\newblock On the expressive power of {D}atalog: tools and a case study.
\newblock {\em Journal of Computer and System Sciences}, 51(1):110--134, 1995.

\bibitem{kv00-conjunctive}
Ph.G. Kolaitis and M.Y. Vardi.
\newblock Conjunctive-query containment and constraint satisfaction.
\newblock {\em Journal of Computer and System Sciences}, 61:302--332, 2000.

\bibitem{kv00-game}
Ph.G. Kolaitis and M.Y. Vardi.
\newblock A game-theoretic approach to constraint satisfaction.
\newblock In {\em Proceedings 17th National (US) Conference on Artificial
  Intellignece, {AAAI'00}}, pages 175--181, 2000.

\bibitem{schaefer78}
Thomas~J. Schaefer.
\newblock The complexity of satisfiability problems.
\newblock In {\em Proceedings of the ACM Symposium on Theory of Computing
  (STOC)}, pages 216--226, 1978.

\end{thebibliography}

\newpage

\appendix

\section{Proof of Theorem \ref{thm:main}}

\begin{proof}
$(\Rightarrow)$
\newcommand{\pr}{\mathsf{pr}}
Let $H$ be a winning strategy for the Duplicator in the $k$-cover game 
on $\rela$ and $\relb$,
let $\relt$ be any structure of coverwidth $\leq k$, let $f$ be any 
homomorphism from $\relt$ to $\rela$,
let $G=(T,F)$ be a scheme for $\relt$ of weight $\leq k$, and
let $v_1,\ldots,v_n$ be a perfect elimination ordering of $G$.

We shall construct a sequence of partial mappings 
$g_0, \ldots, g_n$
from $T$ to $B$ such that for each $i$: 
\begin{enumerate}
\item $\dom(g_i) = \{ v_1, \ldots, v_i \}$, and
\item for every clique $L\subseteq\{v_1,\ldots,v_i\}$ in $G$, 
there exists a projective homomorphism $h \in H$ 
with domain $f(L)$
in the winning strategy of
the Duplicator, such that for every $v\in L$, $h(f(v))=g_i(v)$.
\end{enumerate}

We define $g_0$ to be the partial function with empty domain. For every
$i \geq 0$,
the partial mapping
$g_{i+1}$ is obtained by extending $g_i$ 
in the following way.
As $v_1,\ldots,v_n$ is a perfect elimination ordering, the set 
$$L=\{v_{i+1}\}\cup\{v_j : j<i+1, (v_j,v_{i+1})\in F\}$$
is a clique of $G$.
Define $L'$ as $L\setminus\{v_{i+1}\}$. 
By the induction hypothesis, there 
exists $h \in H$ such that
for every $v\in L'$, $h(f(v))=g_i(v)$.
Let us consider two cases.

If $f(v_{i+1})=f(v_j)$ for
some $v_j\in L'$  then we set $g_{i+1}(v_{i+1})$ to be $g_i(v_j)$. Note 
that in
this case property $(2)$ is satisfied, as every clique in $G$ containing 
$v_{i+1}$ 
is contained in $L$ and $h$ serves as a certificate.
(For any clique not containing $v_{i+1}$, we use the induction hypothesis.)

Otherwise, that is, if $f(v_{i+1})\neq f(v_j)$ for all $v_j\in L'$,  we 
do
the following. First, since the weight of $L$ is bounded above by $k$ 
and $f$
defines an homomorphism from $\relt$ to $\rela$ then the weight of 
$f(L)$ is also bounded by $k$.
Observe that $f(L) = \dom(h)\cup\{f(v_{i+1})\}$. By the
forth property of winning strategy there exists an extension $h' \in H$ of 
$h$ that
is defined over $v_{i+1}$. We set $g_{i+1}(v_{i+1})$ to be 
$h'(f(v_{i+1}))$. Note that $h'$ certifies
that property $(2)$ is satisfied for very clique containing $v_{i+1}$;
again, any clique not containing $v_{i+1}$ is covered by the induction hypothesis.

Finally, let us prove that $g_n$ indeed defines an homomorphism from 
$\relt$ to $\relb$. Let $R$ be any
relation symbol and let $(t_1, \ldots, t_l)$ be any relation in 
$R^{\relt}$. We want to show
that $(g_n(t_1),\ldots,g_n(t_l))$ belongs to $R^{\relb}$. Since 
$G$ is an scheme for $\relt$,
$\{t_1,\ldots,t_l\}$ constitutes a clique of $G$. By property $(2)$ there 
exists $h\in H$
such that $h(f(t_i))=g(t_i)$ for all $i$.
Observing that as $f$ is an homomorphism from $\relt$ to $\rela$, 
we 
can have
that $(f(t_1),\ldots,f(t_l))$ belongs to $R^{\rela}$.
Finally, as $h$ is a projective homomorphism
from $\rela$ to $\relb$, the tuple 
$(h(f(t_1)),\ldots,h(f(t_l)))$ 
must be in $\relb$.

$(\Leftarrow)$
We shall construct a winning strategy $H$ for the Duplicator. We need a 
few definitons.
Fix a sequence $a_1,\dots,a_m$ of elements of $A$.
A {\em valid tuple} for $a_1,\dots,a_m$ is any tuple 
$(\relt,G,v_1,\dots,v_m,f)$ where
$\relt$ is a relational structure, $G$ is an scheme of weight $k$ for 
$\relt$, $\{v_1,\dots,v_m\}$ is a clique 
of $G$, and $f$ is an homomorphism from 
$\relt,v_1,\dots,v_m$ to $\rela,a_1,\dots,a_m$.
(By a homomorphism from $\relt,v_1,\dots,v_m$ to $\rela,a_1,\dots,a_m$,
we mean a homomorphism from $\relt$ to $\rela$ that maps
$v_i$ to $a_i$ for all $i$.)
By $S(\relt,G,v_1,\dots,v_m,f)$ we denote the set of all mappings $h$ 
with domain $\{a_1,\dots,a_m\}$ such that there
is an homomorphism from $\relt,v_1,\dots,v_m$ to 
$\relb,h(a_1),\dots,h(a_m)$. We are now in a situation to define $H$.
$H$ contains for every subset $a_1,\dots,a_m$ of weight at most $k$, 
every partial mapping
$h$ that is contained in all $S(\relt,G,v_1,\dots,v_m,f)$ where 
$(\relt,G,v_1,\dots,v_m,f)$
is a valid tuple for $a_1,\dots,a_m$.

Let us show that $H$ is indeed a winning strategy. First, observe that 
$H$ is nonempty, as it contains the partial function with empty domain.
Second, let us show that $H$ contains only projective homomorphisms. 
Indeed, let $h$ be any mapping in $H$
with domain $a_1,\dots,a_m$, let $R$ be any relation symbol and let 
$(c_1,\dots,c_l)$ be any
tuple in $R^{\rela}$. Let us define $\relt$ to be the substructure (not 
necessarily induced) of $\rela$ with universe
$\{a_1,\dots,a_k,c_1,\dots,c_l\}$ containing only the tuple 
$(c_1,\dots,c_l)$ in $R^{\relt}$.
It  is easy to verify that
the graph $G=(\{a_1,\dots,a_k,c_1,\dots,c_l\},F)$ where 
$F=\{(a_i,a_j):  i\neq j \}\cup\{(c_i,c_j):  i\neq j \}$
is an scheme of $\relt$ of weight $\leq k$. Consequently, 
$(\relt,G,a_1,\dots,a_m,id)$
is a valid tuple for $a_1,\dots,a_m$ and therefore there exists an 
homomorphism $g$ from
$\relt$ to $\relb$, and hence satisfying $(g(c_1),\dots,g(c_l))\in 
R^{\relb}$, such that $g(a_i)=h(a_i)$ for all $i=1,\dots k$.

To show that $H$ is closed under subfunctions is rather easy.
Indeed, let $h'$ be any mapping in $H$
with domain $a_1\dots,a_m$. We shall see that the restriction $h$ of 
$h'$ to $\{a_1,\dots,a_{m-1}\}$
is also in $H$. Let $(\relt,G,v_1,\dots,v_{m-1},f)$ be any valid tuple 
for $a_1,\dots,a_{k-1}$.
We construct a valid tuple $(\relt',G',v_1,\dots,v_m,f')$ for 
$a_1,\dots,a_m$ in the
following way: $v_m$ is a new (not in the universe of $\relt)$ element,
$\relt'$ is the structure obtained from $\relt$
by adding $v_m$ to the universe of $\relt$ and keeping the
same relations, $f'$ is the extension of $f$ in which $v_m$ is map to 
$a_m$,
and $G'$ is the scheme of $\relt$ obtained by adding to $G$ an edge 
$(v_j,v_m)$ for every $j=1,\dots,m-1$.
Since $(\relt',G',v_1,\dots,v_m,f')$ is a valid tuple for 
$a_1,\dots,a_m$ and $h'\in H$, there exists an homomorphism
$g'$ from $\relt',v_1,\dots,v_m$ to $\relb,h'(a_1),\dots,h'(a_m)$. 
Observe then that
the restriction $g$ of $g'$ to $\{a_1,\dots,a_{m-1} \}$ defines then an 
homomorphism from $\relt,v_1,\dots,v_{m-1}$ to 
$\relb,h(a_1),\dots,h(m_1)$.

Finally, we shall show that $H$ has the forth property. The proof 
relies 
in the following easy properties
of the valid tuples. Let $a_1,\dots,a_m$ be elements of $A$ and let 
$(\relt_1,G_1,v_1,\dots,v_m,f_1)$
and let $(\relt_2,G_2,v_1,\dots,v_m,f_2)$ be valid tuples for 
$a_1,\dots,a_m$ such that $T_1\cap T_2=\{v_1,\dots,v_m\}$,
let $\relt$ be $\relt_1\cup\relt_2$
(that is, the structure $\relt$ whose universe is the union of
the universes of $\relt_1$ and $\relt_2$, and in which
$R^{\relt} = R^{\relt_1} \cup R^{\relt_2}$ for all relation symbols $R$),
$G=G_1\cup G_2$ 
and
let $f$ be the mapping from the universe $T$ of $\relt$ to $B$ that 
sets 
$a$ to $f_1(a)$ if $a\in T_1$
and to $f_2(a)$ if $a\in T_2$ (observe that $f_1$ and $f_2$ coincide 
over $\{v_1,\dots,v_m\}$).
Then $(\relt,G,v_1,\dots,v_m,f)$ is a valid tuple for $a_1,\dots,a_m$. 
We call $(\relt,G,v_1,\dots,v_m,f)$,
the {\em union} of $(\relt_1,G_1,v_1,\dots,v_m,f_1)$ and 
$(\relt_2,G_2,v_1,\dots,v_m,f_2)$.
Furthermore,
$S(\relt,G,v_1,\dots,v_m,f)\subseteq 
S(\relt_1,G_1,v_1,\dots,v_m,f_1)\cap S(\relt_2,G_2,v_1,\dots,v_m,f_2)$
(in fact, 
$S(\relt,G,v_1,\dots,v_m,f)=S(\relt_1,G_1,v_1,\dots,v_m,f_1)\cap 
S(\relt_2,G_2,v_1,\dots,v_m,f_2)$,
although we do not need the equality in our proof).

Let $h$ be any mapping in $H$, let
$\{a_1,\dots,a_{m-1}\}$ be its domain,
and let
$a_m$ be any element in the universe of $\rela$ such that 
$\{a_1,\dots,a_m\}$ has weight $\leq k$.
Let us assume, towards a contradiction, that
there is not extension $h'$ of $h$ in $\mathcal H$.  Then there exists 
a {\em finite}
collection $\{(\relt_i,G_i,v_1,\dots,v_m,f_i) : i\in I\}$ of valid 
tuples for $a_1,\dots,a_m$ such
that the intersection $\bigcap_{i\in I} 
S(\relt_i,G_i,v_1,\dots,v_m,f_i)$ does not contain any extension
of $h$. We can rename the elements of the universes so that for every 
different $i,j\in I$ we have that
$T_i\cap T_j=\{v_1,\dots,v_m\}$.

Let $(\relt,G,v_1,\dots,v_m,f)$ be the union of 
$(\relt_i,G_i,v_1,\dots,v_m,f_i)$, $i\in I$,
which is a valid tuple for $a_1,\dots,a_m$. Since 
$$S(\relt,G,v_1,\dots,v_m,f)\subseteq \bigcap_{i\in I} 
S(\relt_i,G_i,v_1,\dots,v_m,f_i)$$
we can conclude that $S(\relt,G,v_1,\dots,v_m,f)$ does not contain any 
extension of $h$. We are almost
at home. It is only necessary to observe that 
$(\relt,G,v_1,\dots,v_{m-1},f)$ is a valid tuple for 
$a_1,\dots,a_{m-1}$
and since $S(\relt,G,v_1,\dots,v_m,f)$ does not contain any extension 
of 
$h$, $S(\relt,G,v_1,\dots,v_{m-1},f)$ cannot
contain $h$, in contradiction with $h\in H$.
\qed \end{proof}

\end{document}